\documentclass{PoS}

\usepackage{xspace}
\usepackage{array}
\usepackage{makecell}
\usepackage{cite}
\usepackage{multirow}
\usepackage[authoryear,square]{natbib}
\bibpunct{(}{)}{;}{a}{}{,}

\title{The SKA as a Doorway to Angular Momentum}

\ShortTitle{Angular Momentum}

\author{\speaker{D. Obreschkow}$^1$, M. Meyer$^1$, A. Popping$^1$, C. Power$^1$, P. Quinn$^1$, L. Staveley-Smith$^1$
\\ 
$^1$International Centre for Radio Astronomy Research (ICRAR)
\\
E-mail: \email{danail.obreschkow@icrar.org}
}

\abstract{Angular momentum is one of the most fundamental physical quantities governing galactic evolution. Differences in the colours, morphologies, star formation rates and gas fractions amongst galaxies of equal stellar/baryon mass~$M$ are potentially widely explained by variations in their specific stellar/baryon angular momentum~$j$. The enormous potential of angular momentum science is only just being realised, thanks to the emergence of the first simulations of galaxies with converged spins, paralleled by a dramatic increase in kinematic observations. Such observations are still challenged by the fact that most of the stellar/baryon angular momentum resides at large radii. In fact, the radius that maximally contributes to the angular momentum of an exponential disk ($3\Re-4\Re$) is twice as large as the radius that maximally contributes to the disk mass; thus converged measurements of angular momentum require either extremely deep IFS data or, alternatively, kinematic measurements of neutral atomic hydrogen (\ha), which naturally resides at the large disk radii that dominate the angular momentum. The SKA has a unique opportunity to become the world-leading facility for angular momentum studies due to its ability to measure the resolved and/or global HI kinematics in very large and well-characterised galaxy samples. These measurements will allow, for example, (1) a very robust determination of the two-dimensional distribution of galaxies in the $(M,j)$-plane, (2) the largest, systematic measurement of the relationship between $M$, $j$, and tertiary galaxy properties, and (3) the most accurate measurement of the large-scale distribution and environmental dependence of angular momentum vectors, both in terms of norm and orientation. All these measurements will represent exquisite tools to build a next generation of galaxy evolution models.}

\FullConference{
Advancing Astrophysics with the Square Kilometre Array\\
June 8-13, 2014\\
Giardini Naxos, Sicily, Italy}

\newcolumntype{L}[1]{>{\raggedright\let\newline\\\arraybackslash\hspace{0pt}}m{#1}}
\newcolumntype{C}[1]{>{\centering\let\newline\\\arraybackslash\hspace{0pt}}m{#1}}
\newcolumntype{R}[1]{>{\raggedleft\let\newline\\\arraybackslash\hspace{0pt}}m{#1}}

\newcommand{\be}{\begin{equation}}
\newcommand{\ee}{\end{equation}}
\newcommand{\eq}[1]{Equation~(\ref{eq_#1})}
\newcommand{\tab}[1]{Table~\ref{tab_#1}}
\newcommand{\fig}[1]{Figure~\ref{fig_#1}}

\newcommand{\kms}{{\rm km\,s^{-1}}}

\newcommand{\ha}{H{\sc\,i}\xspace}
\newcommand{\itha}{H{\footnotesize\,I}\xspace}
\newcommand{\hm}{H$_2$}

\newcommand{\Rs}{R_{\ast}}

\newcommand{\Rflat}{R_{\rm flat}}
\newcommand{\vflat}{V}

\newcommand{\abs}[1]{|#1|}
\renewcommand{\Re}{R_{\rm e}}\newcommand{\half}{\frac{1}{2}}
\newcommand{\mnras}{MNRAS} 
\newcommand{\nat}{Nat} 
\newcommand{\aj}{AJ} 
\newcommand{\apj}{ApJ} 
\newcommand{\apjs}{ApJS} 
\newcommand{\etal}{\textit{et al}}

\begin{document}

\section{Introduction}

\subsection{Importance of angular momentum}

\begin{figure}[b]
	\centering
	\vspace{0mm}
	\includegraphics[width=\textwidth]{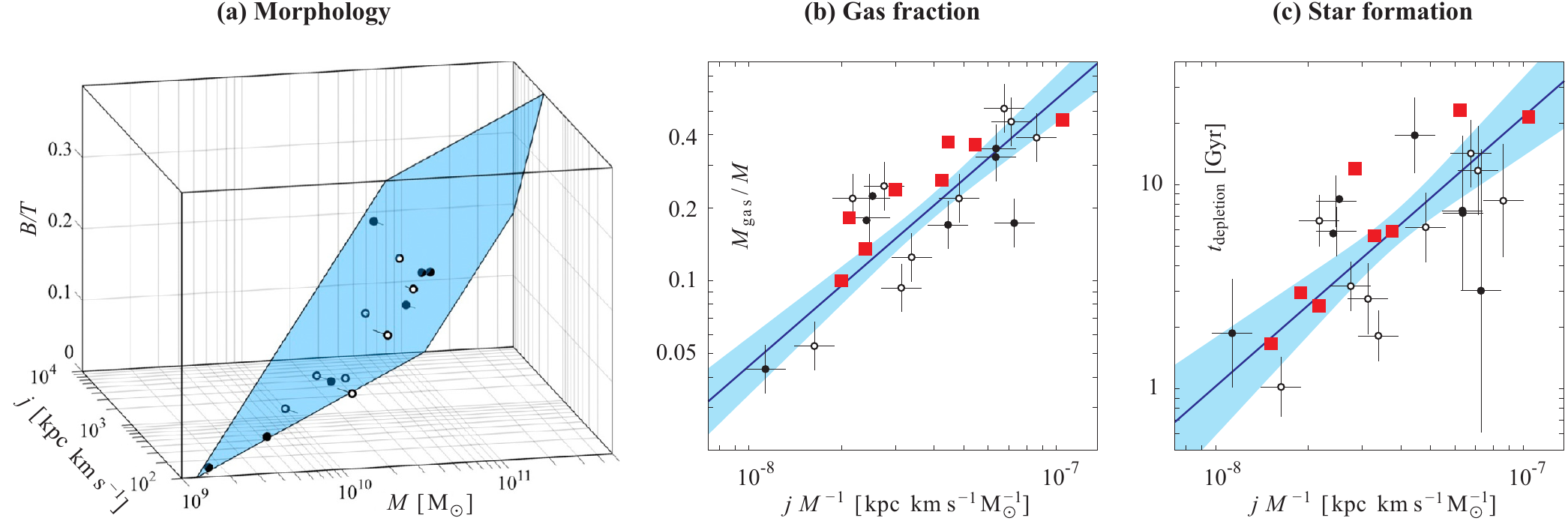}
	\vspace{-3mm}
	\caption{\small [from \cite{Obreschkow2014a} and forthcoming work] Empirical evidence for the fundamental importance of the specific baryon angular momentum $j$, measured 10-times more accurately than so far, using kinematic maps of atomic hydrogen \citep{Leroy2008}. These high-precision data uncover strong correlations between $j$, baryon mass $M$ and tertiary galaxy properties: (a) shows the $M$-$j$-morphology relation, where morphology is measured by the bulge mass fraction $B/T$; (b) and (c) show the relations between $j/M$ and the cold gas fraction and depletion time (gas mass per star-formation rate), respectively. Black dots denote measurements for barred (open dots) and unbarred (filled dots) spiral galaxies. The blue plane in panel (a) and lines in panels (b) and (c) are regressions to the data, with the blue shades in panels (b) and (c) representing the uncertainty of these fits. Red squares are theoretical predictions from high-resolution hydrodynamic simulations [courtesy of C.~Brook].}
	\label{fig_j_obs}
\end{figure}

Understanding how galaxies form and evolve is one of the central topics in explaining the Universe we observe today. Pioneering studies \citep{Fall1980,Quinn1988,Mo1998} have long stressed that galaxy properties are primarily driven by two physical quantities: mass and angular momentum (spin). Two recent advances have significantly increased our ability to understand the role of angular momentum. First, increased computing power and enhanced modelling of stellar feedback finally allow the simulation of galaxies with realistic angular momenta \citep{Governato2010,Agertz2011,Guedes2011,Brooks2011}. Second, enormous progress in kinematic observations, via optical integral field spectroscopy (IFS, see \citealp{Glazebrook2013}) and radio/millimetre interferometry (e.g.\ using the VLA and ALMA), now yield the first precision-measurements of spin in controlled galaxy samples.

The new fields of research now accessible via angular momentum studies range from sub-galactic astrophysics (i.e., the study of angular momentum feedback, spin alignment of individual components, warped disks) to global galaxy evolution studies and large-scale cosmology. The next two paragraphs discuss two fruitful examples that are by no means meant to be exhaustive.

\begin{figure*}[t]
	\centering
	\includegraphics[width=0.93\textwidth]{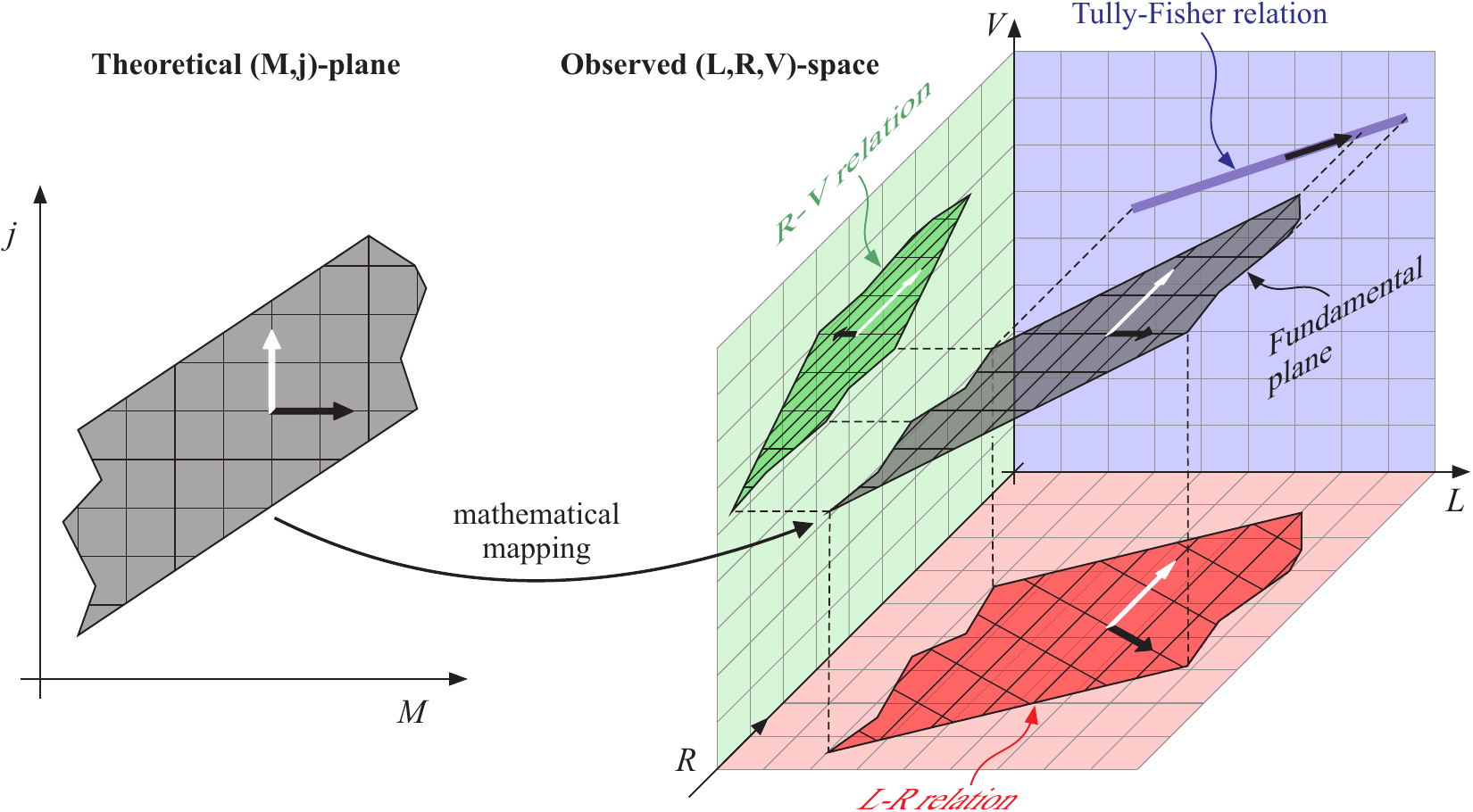}
	\vspace{-1mm}
	\caption{\small [\cite{Obreschkow2014a}] In the model of an exponential disk inside a spherical halo, galaxies lie on a plane in the space spanned by luminosity $L$, disk radius $R$, and rotation velocity $V$. This plane can be explained as a mapping of the ($M$,$j$)-plane into ($L$,$R$,$V$)-space. Projections onto the planes $(L,R)$, $(R,V)$, and $(V,L)$ give rise to classical scaling relations, whose dependence on morphology is then explained by the $M$-$j$-morphology relation.}
	\label{fig_scalings}
\end{figure*}

In the context of galaxy evolution studies, a major result of recent kinematic observations is that the shape of galaxies is tightly linked to their angular momentum and that the historical classification of galaxies by stellar or baryon mass $M$ and Hubble type can instead be substituted for a more physically motivated classification by $M$ and baryon/stellar angular momentum $J$ \citep{Cappellari2011,Romanowsky2012,Obreschkow2014a}. This is illustrated in Fig.~\ref{fig_j_obs}a for spiral galaxies, where we have introduced the {\it specific} angular momentum $j\equiv J/M$ (not to be confused with variations of the spin parameter $\lambda$, also sometimes called `specific angular momentum'). Moreover, theoretical models reveal that the most important scaling relations of spiral galaxies, the mass-size-velocity scalings, can be understood as mappings of the (\textit{M,j})-plane into three dimensions (Fig.~\ref{fig_scalings}). Finally, angular momentum must play a fundamental role in star formation, since it sets the disk pressure, regulating the conversion of atomic gas to molecular gas and stars. The preliminary analysis in Figs.~\ref{fig_j_obs}b and \ref{fig_j_obs}c confirms that in star-forming galaxies $j/M$ is significantly correlated with the cold gas fraction and gas depletion time. This correlation had to be expected, to first order, because $j/M=J/M^2$ is inversely proportional to the disk surface density (details in \citeauthor{Obreschkow2014a}), which affects the conversion rate of \ha to \hm\ and stars.

\begin{figure}[t]
	\centering
	\vspace{-2mm}
	\includegraphics[width=11.7cm]{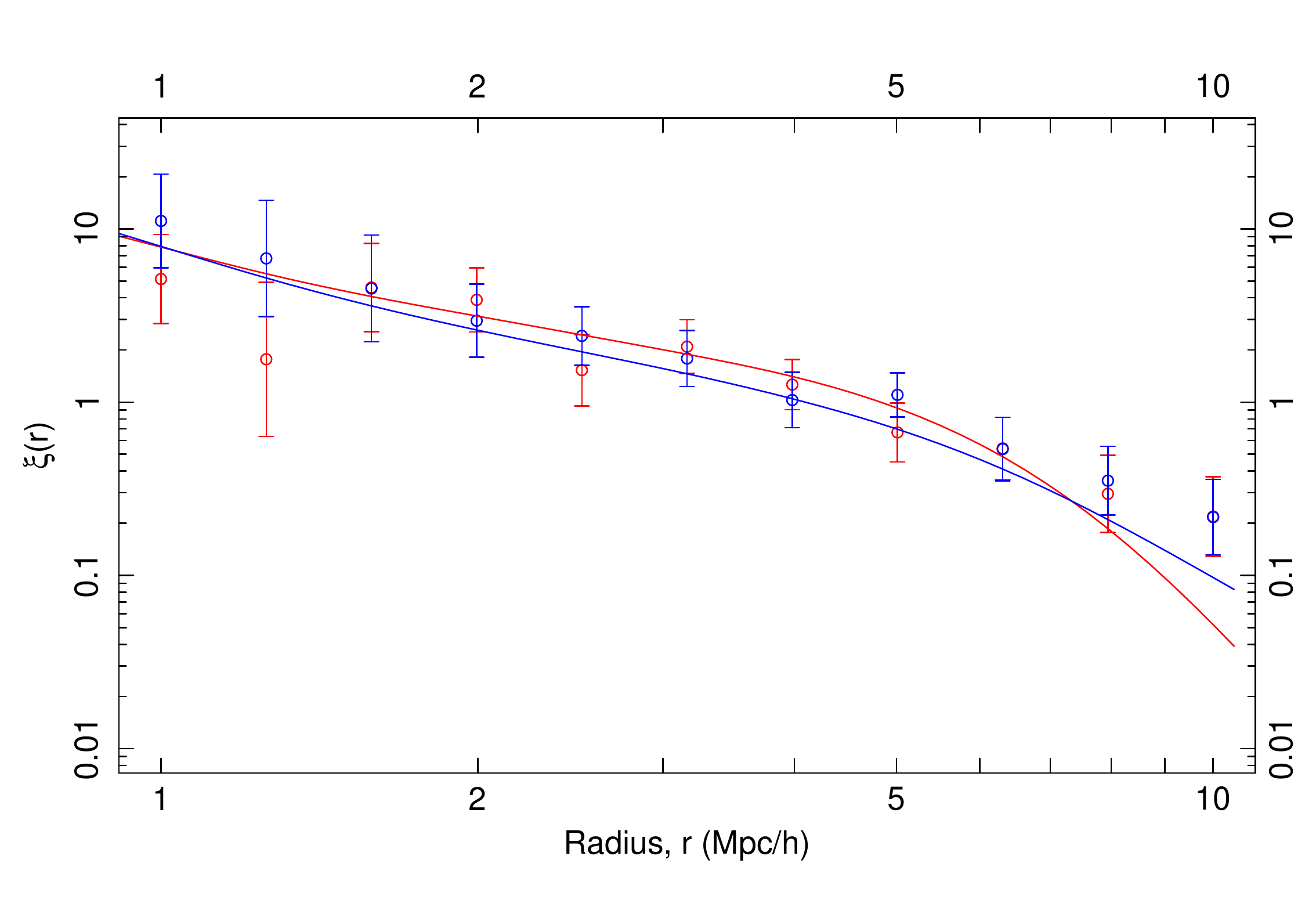}
	\vspace{-3mm}
	\caption{\small Preliminary analysis of the 2-point correlation function of spiral galaxies split in two samples of low (red) and high (blue) specific angular momentum $j$, relative to the mean $M$-$j$ relation. Points with (correlated) error bars are observational data from the HIPASS catalog with optical counterparts from SuperCOSMOS \citep{Doyle2006}. Lines are model results from the S$^3$-SAX model \citep{Obreschkow2009b}. Observations and simulation are consistent within the observational uncertainties, but the latter are currently too large to verify the predicted enhanced clustering of low-$j$ galaxies.}
	\label{fig_correlation}
\end{figure}

In a cosmological context, much progress is to be made by analysing the large-scale ($>$Mpc, comoving) distribution of galaxies as a function of kinematic properties, in simulated and observed datasets. In fact, the large-scale distribution of angular momentum, both in terms of norm and orientation, contains information on the mechanisms of angular momentum growth by cosmic tides \citep{White1984} and the redistribution of angular momentum by galaxy coalescence. Empirical investigations using the SDSS survey revealed that spiral and elliptical galaxies obey opposite correlations with filaments \citep{Jones2010}, hinting at the importance of the orbital angular momentum involved in the merger-based formation of ellipticals. As for the norm of $j$, Fig.~\ref{fig_correlation} shows a state-of-the-art analysis of the galaxy-galaxy correlation function for two samples of low and high $j$, relative to the mean $M$-$j$ relation. Although the models (lines) appear consistent with the observations (points) within the observational uncertainties, the latter are currently too large to verify the predicted enhanced clustering of the low-$j$, subsample.

\subsection{Objective of this paper}

Having illustrated some of the importance of angular momentum in galaxies (Section~1.1), the remaining objective of this paper is to demonstrate the world-leading potential of the SKA in this field due to its unprecedented ability to detect neutral atomic hydrogen (\ha) via 21cm interferometry. In fact, kinematic maps of \ha combined with optical images allow the stellar/baryon $j$ of disk galaxies to be measured within $\lesssim\!5\%$ \citep{Obreschkow2014a} -- the most accurate measurements to-date. \emph{The advantage of \itha data relies in the fact that \itha naturally resides at large disk radii (about twice as large as stars), such that the radii that maximally contribute to the \itha mass are similar to those that maximally contribute to angular momentum ($3\Re-4\Re$).}

Section~2 explains how both spatially resolved kinematic \ha maps and unresolved \ha line profiles can be used to measure the angular momentum in disk galaxies. Simulation-based modelling in Section~3 reveals that already the SKA1 will increase the numbers of good kinematic \ha~maps and unresolved \ha~detections by about two orders of magnitude relative to the current state-of-the-art (e.g.~THINGS, HIPASS, ALFALFA). Explicit numbers of angular momentum measurements are calculated for different benchmark survey scenarios with well-defined observing parameters. A brief discussion and summary are presented in Section~4.

\section{Methods of Measuring Angular Momentum}\label{section_research}

This chapter presents different ways of measuring the stellar/baryon angular momentum $\mathbf{J}$ of disks using 21cm radio data. We will restrict this discussion to the measurement of the norm $J\equiv|\mathbf{J}|$, however, using the inclination+chirality of the disk (from images or kinematic maps), the norm $J$ can be converted to the vector $\mathbf{J}$. With respect to the centre of mass, $J$ is defined as
\be
	J = \left|\int dM~\mathbf{r}\times\mathbf{v}\,\right|,
\ee
where $\mathbf{r}$ is the position vector of the mass element $dM$ relative to the centre of mass and $\mathbf{v}$ denotes the velocity in an inertial frame. In principle, $J$ can be measured for different baryonic components, including the atomic and molecular material of the disk, but often only the stellar component is considered due to its dominance of the disk angular momentum in nearby galaxies \citep[e.g.][Table~1]{Obreschkow2014a}.

\subsection*{Method 1: Precision Measurement (requiring 21cm maps and resolved optical images)}\label{subsection_method1}

In the approximation of a disk with circular orbits, the norm of the angular momentum relative to the centre of gravity can be rewritten as
\be\label{eq_J_general}
	J = 2\pi \int_0^\infty dr\,r^2\,\Sigma(r)\,v(r)
\ee
and the specific angular momentum as
\be\label{eq_j}
	j \equiv \frac{J}{M} = \frac{\int_0^\infty dr~r^2~\Sigma(r)~v(r)}{\int_0^\infty dr~r~\Sigma(r)},
\ee
where $v(r)$ is the norm of the circular velocity at $r=\abs{\mathbf{r}}$, and $\Sigma(r)$ is the azimuthally averaged mass surface density of the considered baryonic component. Note that $r$, $v$ and $\Sigma$ have to be deprojected (i.e., reconstructed from observations of an \textit{inclined} disk) assuming circular orbits. The use of axially averaged surface densities $\Sigma(r)$ has the advantage of increasing the signal-to-noise in the faint outer part of the disk. This axial averaging, does \textit{not}, in fact, lead to errors if the  surface density were not axially symmetric, as long as the velocity field is axially symmetric.

\begin{figure}[t]
	\centering
	\vspace{-2mm}
	\includegraphics[width=10cm]{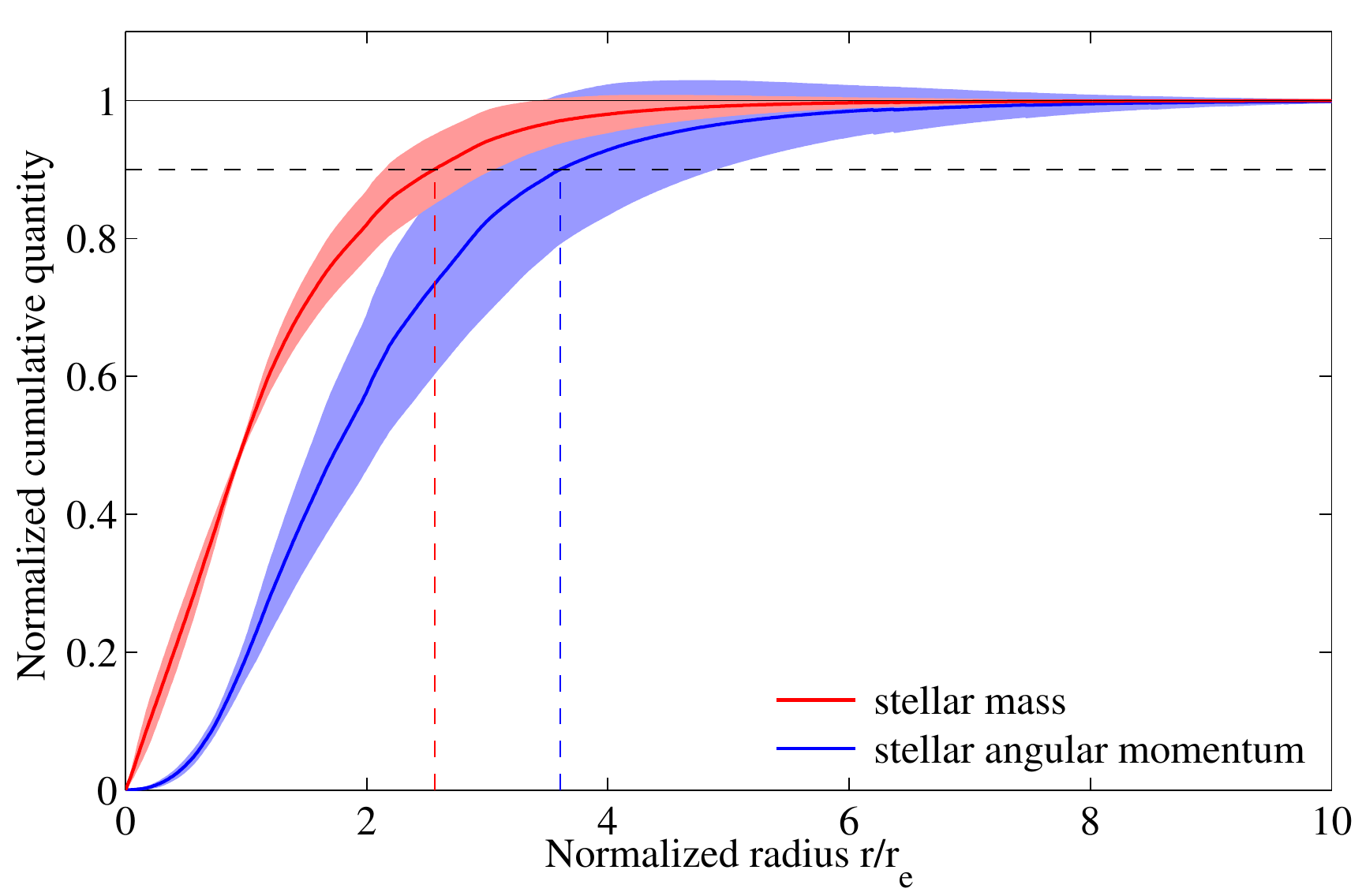}
	\vspace{-4mm}
	\caption{\small Cumulative mass (red) and angular momentum (blue) as a function of the normalised radius for 16 nearby disk galaxies of the THINGS survey (see \citealp{Obreschkow2014a} for details).}
	\label{fig_convergence}
\end{figure}

In practice, the surface density $\Sigma(r)$ is measured from optical images (for stellar $\Sigma$) or from more complex multi-wavelength images (for baryon $\Sigma$). In turn, $v(r)$ can be measured from kinematic maps. In 16 regular disk galaxies of the THINGS survey \citep{Leroy2008}, axially averaged data of the stellar and baryon $\Sigma(r)$ (from optical+CO+\ha imaging) and $v(r)$ (from \ha maps) is available at sub-kpc resolution out to $\sim\!14$ exponential disk scales. This extremely deep data enabled us to study the convergence of the stellar and baryon $J$ as a function of $r$ in detail (see Fig.~\ref{fig_convergence}). For these 16 galaxies $J$ and $j$ are converged to much less than $1\%$, reducing the overall uncertainties of $j$ to 3\%-5\%.

 \fig{convergence} highlights the key challenge of such precision measurements of angular momentum: the exquisite surface brightness sensitivity requirement. The data reveals that local disk galaxies need to be kinematically mapped to galactocentric radii of $3.6\Re$, on average, in order to capture 90\% of the stellar angular momentum. For IFS surveys, this is an extreme requirement, given that the majority of galaxies in current IFS surveys are mapped to less than $2\Re$. Pushing from $2\Re$ to $3.6\Re$ corresponds to an increase of the surface brightness sensitivity by $\sim3$~magnitudes, i.e.~a $\sim200$-times increase in observing time. In turn, \ha~maps almost by default extend to at least $4\Re$, because the \ha gas is naturally situated at larger radii than the stars.

\subsection*{Method 2: Approximation (requiring global 21cm profiles and optical images)}
If \ha kinematic maps are unavailable, the simplest assumption to use is that galactic disks are objects rotating at constant velocity $V$ with exponentially declining surface density of scale length $R$. In this model, the specific angular momentum is exactly $j=2RV$ according to \eq{j}. In practice, the scale radius $\Rs\approx\Re/1.68$ of the stellar disk can be estimated from optical images, and for non-dwarf galaxies the asymptotic circular velocity (now considered the sole velocity) is well approximated as half the \ha linewidth $W_{50}$, corrected for the inclination $i$. Thus,
\be\label{eq_j1}
	j = \Rs~W_{50}~\sin^{-1}i.
\ee
A detailed comparison of this approximation against high-precision measurements of the stellar $j$ from the sub-kpc kinematic maps of the THINGS sample revealed that Eq.~(\ref{eq_j1}) recovers the `true' $j$ with an observational error of about 40\% (standard deviation, see Section 2.4 in \citealp{Obreschkow2014a}). \cite{Romanowsky2012} demonstrated that this is just good enough to evidence a systematic angular momentum dependence of morphology within regular late-type galaxies. However, at this stage it is not yet clear to what extent this conclusion holds for galaxies with poorly defined or inexistent asymptotic velocities.

\subsection*{Method 3: Approximation (requiring global 21cm profiles and IFS velocity maps)}
The biggest drawback of the approximation method 2 is that the errors of $j$ are correlated with the stellar/baryon mass $M$ as a result of the rotation curve shape being correlated to $M$ \citep{DeBlok2008}. To bypass this limitation, the assumption of a constant rotation velocity can be substituted for a variable velocity model $v(r)=\vflat[1-\exp(-r/R_{\rm flat})]$ \citep{Boissier2003}, where $\vflat$ now denotes the asymptotic velocity at large radii $r$, and $\Rflat$ is the scale-length of the rotation curve. Since most \ha is generally situated at radii much larger than $\Rflat$, the velocity $\vflat$ remains well approximated by $\vflat=\frac{1}{2}W_{50}~\sin^{-1}i~$ in this model. Convolving $v(r)$ with an exponential surface density $\Sigma(r)\propto\exp(-r/\Rs)$ in \eq{j}, results in
\be\label{eq_j2}
	j = \Rs~W_{50}~\sin^{-1}i~\frac{(1+x)^3-1}{(1+x)^3}.
\ee
This equation differs from \eq{j1} by the right-most factor that depends on $x\equiv\Rs/\Rflat$. Thus, this method requires a measurement of $\Rflat$, representing the characteristic scale of the inner part of the rotation curve. Velocity maps from optical IFS data are well-suited for this measurement, even if limited to radii of $1\Re-2\Re$. \cite{Obreschkow2014a} showed that \eq{j2} increases the accuracy relative to \eq{j1} by a factor $2-3$, to a relative error on $j$ of $\sim20\%$ with no significant correlation to $M$. Like in method 2, it is not yet fully established to what extent this method works for galaxies with poorly defined or inexistent asymptotic velocities, such as dwarfs.

\section{Assessment of SKA Capabilities}\label{section_capabilities}

\subsection{Modelling of galaxies and detection rates}

For the following performance calculations we draw a 100 deg$^2$ mock cone from the S$^3$-SAX simulation \citep{Obreschkow2009b,Obreschkow2009f}, truncated radially to redshifts below $z=1.2$, the distance limit for \ha set by the 650~MHz lower limit of SKA1-SUR band~2. This simulation is publicly available and has been verified against local observations in terms of the \ha mass function \citep{Obreschkow2009b}, the \ha velocity function \citep{Obreschkow2013a}, the Tully-Fisher relation \citep{Obreschkow2009b}, and the \ha large-scale clustering (Fig.~\ref{fig_correlation}).\\

A galaxy in the mock cone is considered detected sufficiently well to measure linewidths $W_{50}$ if its peak flux density $s_{\rm peak}$ -- the flux density at the horns of a typical \ha profile -- exceeds 5-times the RMS channel noise ($10\,\kms$ channels in observer-frame). Out of all the galaxies detected in this sense, we determine the subsample for which a GAMA-like survey can detect and spatially resolve the optical counterparts to measure the disk radius and inclination. Explicitly, this subsample is defined by the additional selection criteria $m_{\rm R}<20$ and $R_{\rm e}>1^{\prime\prime}$. These are the minimal criteria required for an approximate angular momentum measurement via `Method 2'. Tables \ref{tab_ska1}--\ref{tab_ska2} show the numbers of galaxies in this subsample, as well as their median redshift.\\

Of particular interest for angular momentum science are those galaxies with kinematic \ha maps that are good enough to allow precision measurements of angular momentum via Method~1. Here, we say that a galaxy has a `good kinematic map', if two criteria are met:
\begin{enumerate}
\item The \ha-half-mass radius spans more than 5 synthesised beams, i.e.~the \ha disk is resolved in $\gtrsim\!100$ beams. (Velocity measurements in 5 galactocentric rings are roughly the minimum requirement for a robust fit of kinematic models.)
\item The average flux per beam inside the \ha-half-mass radius exceeds 10-times the RMS beam noise ($10\,\kms$ channels). (This condition approximately corresponds to a 5-sigma column sensitivity limit of $2\cdot10^{20}\rm\,cm^{-2}$, although the exact value depends slightly on the \ha surface density profiles of the galaxies detected in the respective survey scenarios.)
\end{enumerate}

\begin{figure*}[t]
	\centering\vspace{-5mm}
	\begin{tabular}{cc}
		\hspace{-5mm}\includegraphics[width=8.1cm]{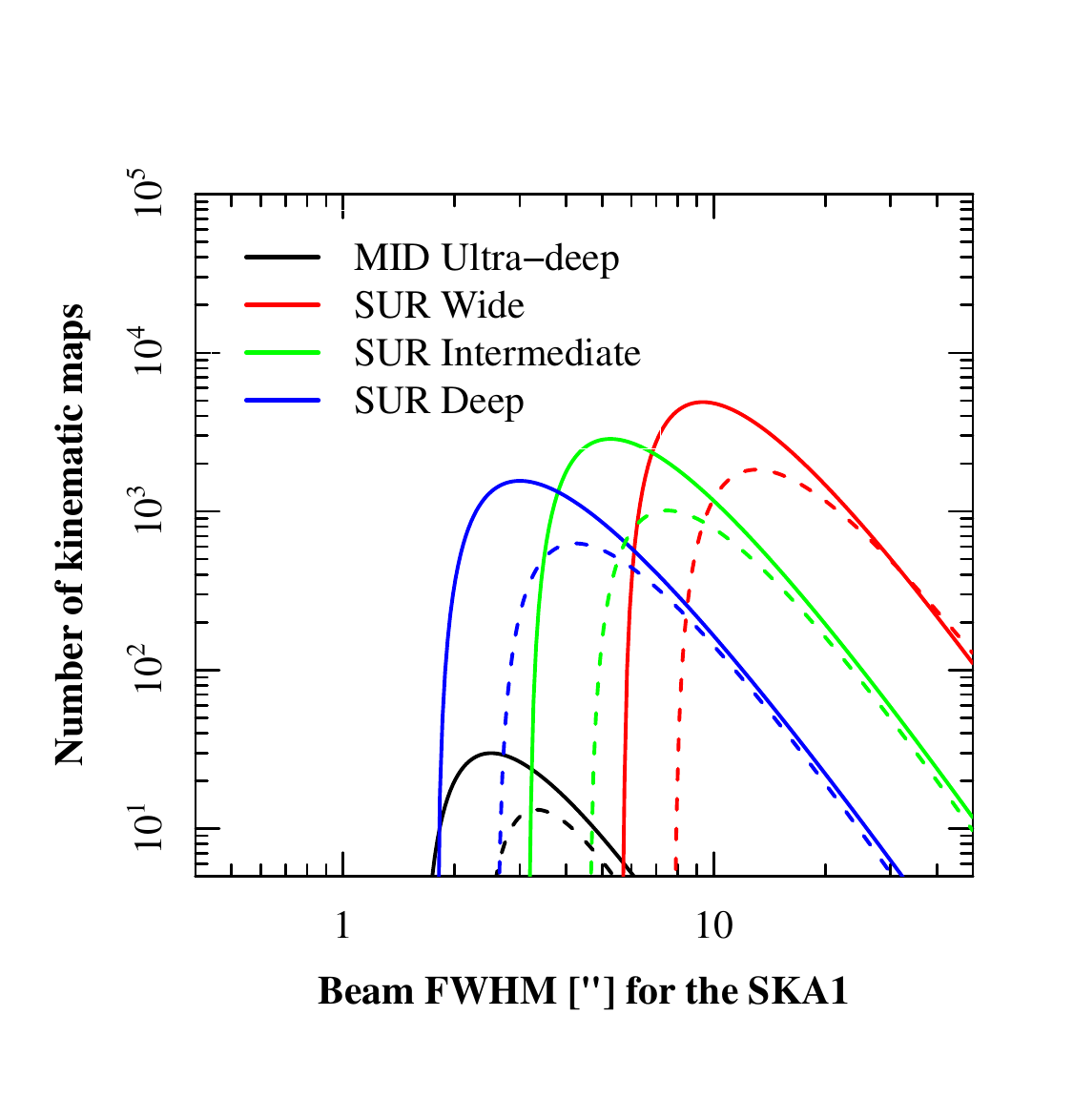} & \hspace{-7mm}\includegraphics[width=8.1cm]{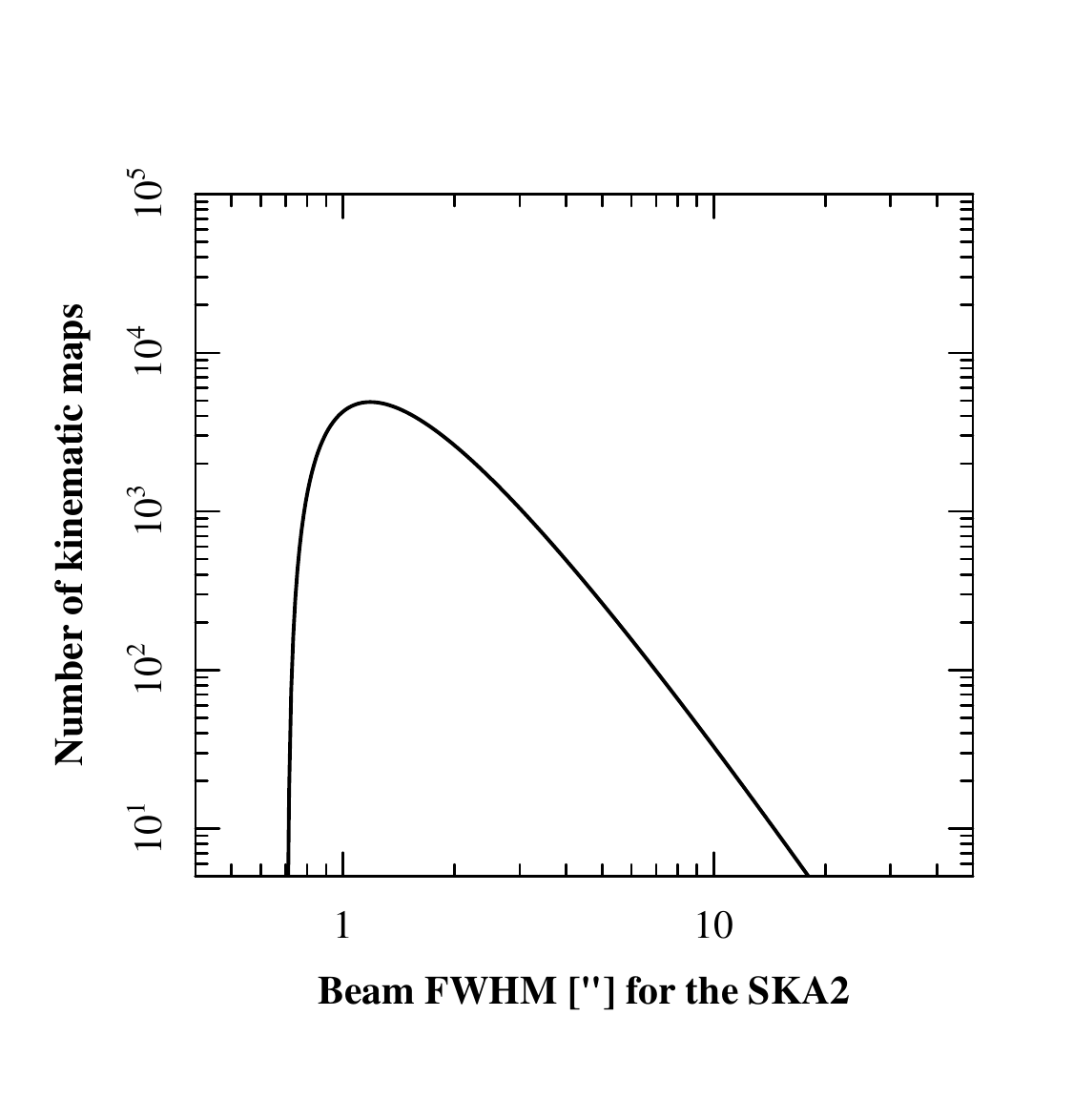}\vspace{-7mm}
	\end{tabular}
	\caption{Number of galaxies with good kinematic \ha maps (as defined in \S3.1), as a function of the FWHM of the synthesised beam (pixel). The left panel represents the SKA1 (solid lines) and the Early phase 50\% SKA1 (dashed lines); the right panel represent the SKA2. The competing criteria that galaxies must be spatially resolved, but still yield enough signal per pixel causes these functions to peak. The increased sensitivity of the SKA2 allows kinematic maps to be obtained for galaxies at greater distances, hence requiring smaller pixels. All these functions assume channels of $10\,\kms$ (observer-frame) with an RMS noise per beam that is independent of the beam FWHM -- an assumption that is approximately verified for the `logarithmic' antenna distribution of the System Baseline Design.}
	\label{fig_fwhm}
\end{figure*}

This definition of a `good kinematic map' ensures a measurement of the stellar/baryon specific angular momentum at less than 10\% uncertainty (down to 3\%-5\% for THINGS-like quality), provided stellar surface density maps from optical imaging. A subtle aspect is the choice of the synthesised beam size (i.e.~the weighting of the baselines). The resolution-criterion (1) favours small beams, whereas the signal/noise-criterion (2) demands large beams. To optimise the case, we have determined the number of good kinematic maps as a function of the beam FWHM (see Fig.~\ref{fig_fwhm}). Table~1 shows the maximally achievable number of good kinematic maps and the corresponding beam FWHM. These FWHM are consistent with those found by \cite{DeBlok2014} for a \ha column sensitivity of $2\cdot10^{20}\rm\,cm^{-2}$ (5-sigma level with $10\,\kms$ channels).

\subsection{SKA1}\label{subsection_ska1}

To predict the capabilities of the SKA1, as specified in the System Baseline Design \citep{Dewdney2013}, we adopt the reference survey scenarios considered at the \ha Science Assessment Workshop (September 2013), but ignore the surveys dedicated to \ha absorption and to the local ISM. Our sensitivity estimations build on a telescope simulation (described in \citealp{Popping2014}), assuming observations at a declination of $-30^\circ$. The respective RMS noise levels per beam for an 8h integration with $10\,\kms$ channels (observer-frame) are approximately 0.1~mJy (SKA1-MID) and 0.25~mJy (SKA1-SUR). These noise levels are reasonably constant ($\sim 50\%$ variations) across the ranges of frequency and beam size relevant to this analysis. The resulting sensitivities of the reference surveys are listed in \tab{ska1}.

\begin{table*}[t]
	\centering
	\normalsize
	\begin{tabular}{L{3.4cm}|C{2.4cm}|C{2.4cm}C{2.4cm}c}
	\hline\hline
	 Telescope     & SKA1-MID & \multicolumn{3}{c}{SKA1-SUR} \\ [0.5ex]
	 \hline
	 \multirow{2}{*}{Survey mode} & Ultra-deep & Wide & Intermediate & Deep \\ [0.5ex]
	 & $2000\rm\,h$,~$3\rm\,deg^2$ & $2\rm\,yr$,~$3\cdot10^4\rm\,deg^2$ & $2\rm\,yr$,~$3\cdot10^3\rm\,deg^2$ & $2\rm\,yr$,~$300\rm\,deg^2$ \\ [1.0ex]
	 \hline
	 FoV $\rm[deg^2]$ & $0.49\cdot(1+z)^2$          & 18                          & 18                          & 18                          \\ [1.5ex]
 Number of pointings&          6             &       1,700            &        170             &         17              \\ [1.5ex]
 Effective~int.~time $\rm[h]$ & $330\cdot(1+z)^2$           & 10                          & 100                         & 1,000                       \\ [1.5ex]
 Beam noise $\sigma$ [mJy] at fixed rest-frame $\Delta V$ of $10~\kms$ & $0.016(1\!+\!z)^{-\half}$   & $0.22(1+z)^{\half}$         & $0.071(1+z)^{\half}$        & $0.022(1+z)^{\half}$        \\ [4.5ex]
 \hline
 Number of line detections, $S_{\rm int}\!>\!5\sigma$&      22,000            &    6,000,000           &    2,500,000           &    1,100,000            \\ [2.5ex]
 Median $z$ & 0.25                        & 0.06                        & 0.11                        & 0.19                        \\ [1.5ex]
 \hline
 Number of line detections, $s_{\rm peak}\!>\!5\sigma$&       9,300            &    2,200,000           &     980,000            &     440,000             \\ [2.5ex]
 Median $z$ & 0.20                        & 0.04                        & 0.09                        & 0.14                        \\ [1.5ex]
 \hline
 ... $s_{\rm peak}\!>\!5\sigma$ and $m_{\rm R}\!<\!20,~R_{\rm e}\!>\!1^{\prime\prime}$&       2,100            &    1,800,000           &     650,000            &     160,000             \\ [2.5ex]
 Median $z$ & 0.19                        & 0.04                        & 0.09                        & 0.16                        \\ [1.5ex]
 \hline
 Number of good kine\-matic maps&         30             &       3,900            &       3,600            &       1,900             \\ [2.5ex]
 Optimal beam FWHM $\rm["]$ & 2.5                         & 9.3                         & 5.3                         & 3.0                         \\ [2.5ex]

	\hline\hline
	\end{tabular}
	\caption{Predicted capabilities of different reference surveys with the SKA1, as specified in the System Baseline Design \citep{Dewdney2013}. The reference survey scenarios are adopted from the \ha Science Assessment Workshop (September 2013). The values chosen for the beam FWHM are the `sweet spots', where the number of kinematic \ha~maps sufficient for a precision-measurement of $j$ is maximal (see Fig.~5). For smaller synthesised beams, the signal per beam becomes so low that most kinematic maps are no longer useable, while larger synthesised beams lead to insufficient spatial resolution.}
	\label{tab_ska1}
\end{table*}

Table~1 confirms that wide-field surveys detect a larger number of sources than deep surveys. The wide-field survey with SKA1-SUR will increase both the number of global \ha line detections and the number of `good kinematic maps' by about two orders of magnitude with respect to the state-of-the-art (e.g.~THINGS, HIPASS, ALFALFA). Explicitly, about 3,900 galaxies will be useable for a precision angular momentum measurement via method 1, while 1,800,000 galaxies will be useable for an approximate angular momentum measurement via method 2 -- provided sufficient optical counterparts.

\newpage
\subsection{Early phase SKA1 (50\% SKA1)}

Table~\ref{tab_halfska1} shows the results for the same analysis as in \S\ref{subsection_ska1}, but assuming half the sensitivity for the SKA1. Halving the sensitivity of the SKA1 reduces the number of useable line detections, as well as the number of good kinematic maps by about a factor two, depending on the survey strategy.\\

\begin{table*}[h!]
	\centering
	\normalsize
	\begin{tabular}{L{3.4cm}|C{2.4cm}|C{2.4cm}C{2.4cm}c}
	\hline\hline
	 Telescope     & 50\%-SKA1-MID & \multicolumn{3}{c}{50\%-SKA1-SUR} \\ [0.5ex]
	 \hline
	 \multirow{2}{*}{Survey mode} & Ultra-deep & Wide & Intermediate & Deep \\ [0.5ex]
	 & $2000\rm\,h$,~$3\rm\,deg^2$ & $2\rm\,yr$,~$3\cdot10^4\rm\,deg^2$ & $2\rm\,yr$,~$3\cdot10^3\rm\,deg^2$ & $2\rm\,yr$,~$300\rm\,deg^2$ \\ [1.0ex]
	 \hline
	 FoV $\rm[deg^2]$ & $0.49\cdot(1+z)^2$          & 18                          & 18                          & 18                          \\ [1.5ex]
 Number of pointings&          6             &       1,700            &        170             &         17              \\ [1.5ex]
 Effective~int.~time $\rm[h]$ & $330\cdot(1+z)^2$           & 10                          & 100                         & 1,000                       \\ [1.5ex]
 Beam noise $\sigma$ [mJy] at fixed rest-frame $\Delta V$ of $10~\kms$ & $0.031(1\!+\!z)^{-\half}$   & $0.45(1+z)^{\half}$         & $0.14(1+z)^{\half}$         & $0.045(1+z)^{\half}$        \\ [4.5ex]
 \hline
 Number of line detections, $S_{\rm int}\!>\!5\sigma$&       9,600            &    2,400,000           &    1,100,000           &     440,000             \\ [2.5ex]
 Median $z$ & 0.19                        & 0.04                        & 0.09                        & 0.13                        \\ [1.5ex]
 \hline
 Number of line detections, $s_{\rm peak}\!>\!5\sigma$&       3,700            &     850,000            &     410,000            &     180,000             \\ [2.5ex]
 Median $z$ & 0.14                        & 0.04                        & 0.06                        & 0.10                        \\ [1.5ex]
 \hline
 ... $s_{\rm peak}\!>\!5\sigma$ and $m_{\rm R}\!<\!20,~R_{\rm e}\!>\!1^{\prime\prime}$&       1,500            &     740,000            &     320,000            &      99,000             \\ [2.5ex]
 Median $z$ & 0.16                        & 0.04                        & 0.06                        & 0.11                        \\ [1.5ex]
 \hline
 Number of good kine\-matic maps&         13             &       2,400            &        930             &        670              \\ [2.5ex]
 Optimal beam FWHM $\rm["]$ & 3.3                         & 13                          & 7.4                         & 4.2                         \\ [2.5ex]

	\hline\hline
	\end{tabular}
	\caption{Same as Table~1, but for the Early phase SKA1 (50\% SKA1).}
	\label{tab_halfska1}
\end{table*}

\newpage\subsection{SKA2}

Finally, we consider the SKA2, consisting of a MID array with 5-times\footnote{For SKA2, a sensitivity increase of a factor of 10 is assumed, but this refers to the natural sensitivity. A sensitivity loss of 50\% has to be applied in addition due to the tapering used (see Braun, 2014, imaging document).} the sensitivity of SKA1-MID (above 350~MHz) and 20-times the instantaneous field-of-view (FoV). This results in a FoV of $9.8\rm~deg^2$ at $z=0$, larger than the $3\rm~deg^2$ of the ultra-deep reference survey for SKA1-MID. Hence, we increase this survey area by a factor 20 to $60\rm~deg^2$. The fact that the SKA2 also allows much higher spatial resolution than SKA1-MID is relevant, since the analysis of the optimal beam size in Fig.~\ref{fig_fwhm} indeed favours SKA2 to work at higher spatial resolution than SKA1-MID to maximise the number of good kinematic \ha~maps.\\

\begin{table*}[h!]
	\centering
	\normalsize
	\begin{tabular}{L{4.4cm}|C{3cm}}
	\hline\hline
	 Telescope     & SKA2 \\ [0.5ex]
	 \hline
	 \multirow{2}{*}{Survey mode} & Ultra-deep \\ [0.5ex]
	 & $2000\rm\,h$,~$60\rm\,deg^2$ \\ [1.0ex]
	 \hline
	 FoV $\rm[deg^2]$ & $9.8\cdot(1+z)^2$           \\ [1.5ex]
 Number of pointings&          6             \\ [1.5ex]
 Effective~int.~time $\rm[h]$ & $330\cdot(1+z)^2$           \\ [1.5ex]
 Beam noise $\sigma$ [mJy] at fixed rest-frame $\Delta V$ of $10~\kms$ & $0.0031(1\!+\!z)^{-\half}$  \\ [4.5ex]
  \hline
 Number of line detections, $S_{\rm int}\!>\!5\sigma$&    2,800,000           \\ [2.5ex]
 Median $z$ & 0.47                        \\ [1.5ex]
  \hline
 Number of line detections, $s_{\rm peak}\!>\!5\sigma$&    1,500,000           \\ [2.5ex]
 Median $z$ & 0.41                        \\ [1.5ex]
  \hline
 ... $s_{\rm peak}\!>\!5\sigma$ and $m_{\rm R}\!<\!20,~R_{\rm e}\!>\!1^{\prime\prime}$&      53,000            \\ [2.5ex]
 Median $z$ & 0.21                        \\ [1.5ex]
  \hline
 Number of good kine\-matic maps&       5,700            \\ [2.5ex]
 Optimal beam FWHM $\rm["]$ & 1.2                         \\ [2.5ex]

	\hline\hline
	\end{tabular}
	\caption{Same as Table~1, but for the SKA2.}
	\label{tab_ska2}
\end{table*}

\newpage
\section{Discussion and Conclusions}\label{section_conclusion}

This paper demonstrates that the SKA will be the \emph{best} machine to conduct world-leading angular momentum science. In fact, kinematic \ha~maps from 21cm interferometry, combined with optical images, allow the specific angular momentum $j$ of disks to be measured within 3\%-5\% uncertainty \citep{Obreschkow2014a} -- the most accurate measurements to-date. It is important to emphasise that \ha~maps outperform optical IFS maps due to the fact that only \ha (not stars) resides where most of the stellar/baryon angular momentum resides. This feature also allows us to measure $j$ to a lesser precision from spatially unresolved \ha~emission lines combined with spatially resolved optical images.

Already the SKA1 will increase the number of kinematic \ha~maps, as well as the number of global \ha line profiles useable for angular momentum measurements by two orders of magnitude compared to the state-of-the-art. It will be challenging to match the millions of unresolved \ha detections with sufficiently many and sufficiently resolved optical counterparts. Our analysis suggests that a seeing-limited ($\sim0.\!''5$ resolution) 3$\pi$-sky imaging+redshift survey at the depth of the GAMA survey provides enough optical counterparts to determine the specific angular momenta of millions of galaxies (using Method 2) and thousands of galaxies (using Method 1).\\

Some science examples enabled by angular momentum measurements with the SKA include
\begin{itemize}
	\item a very robust determination of the two-dimensional distribution in the $(M,j)$-plane of galaxy samples with well-described completeness,
	\item the largest, systematic measurement of the relationship between $M$, $j$, and tertiary galaxy properties (morphology, gas fraction, depletion time),
	\item the most accurate measurement of the large-scale distribution, including environment-depen\-dence, of angular momentum, both in terms of norm and orientation.
\end{itemize}
These examples are by no means exhaustive, but they are good illustrations of powerful future measurements to test and improve hydrodynamical simulations of galaxies and to build a next generation of models of galaxy evolution.

\subsection*{Acknowledgements}

We thank the referees, Erwin de Blok and Paolo Serra, for very insightful and helpful comments.


\end{document}